\documentclass[trackchanges,twocolumn,twocolappendix]{aastex701}
\usepackage{amsmath}

\begin{document}

\title{Dispersive Properties of MHD Waves in the Expanding Solar Wind for a Parker Spiral Geometry}

\author[orcid=0009-0000-5670-0522]{Sebastián Saldivia}
\affiliation{Departamento de Física, Facultad de Ciencias, Universidad de Chile, Las Palmeras 3425, Santiago, 7800003, Chile}
\email[show]{sebastian.saldivia@ug.uchile.cl}

\author[0000-0002-7085-658X]{Felipe Asenjo}
\affiliation{Facultad de Ingeniería y Ciencias, Universidad Adolfo Ibáñez, Santiago 7491169, Chile.}
\email[show]{felipe.asenjo@uai.cl}

\author[0000-0002-9161-0888]{Pablo S. Moya}
\affiliation{Departamento de Física, Facultad de Ciencias, Universidad de Chile, Las Palmeras 3425, Santiago, 7800003, Chile}
\email[show]{\\pablo.moya@uchile.cl}

\begin{abstract}

In this work, we quantify the effects of solar wind expansion on the dispersive properties of the three normal modes of ideal MHD using the Expanding Box Model, under a background magnetic field that follows the Parker spiral geometry. From the linearized MHD-EBM equations, we construct the dispersion tensor and derive analytical expressions for the eigenfrequencies $\omega(k,R)$, magnetic compressibility $C_B$, and the ratio of the parallel electric field to the perpendicular magnetic field  $|\delta E_\parallel|/|\delta B_\perp|$ of the magnetosonic modes to quantify how radial solar wind expansion reshapes the character of compressive fluctuations in the solar wind. Magnetic compressibility increases with heliocentric distance, and this trend shows a better alignment with in-situ observations when expansion is included from the MHD-EBM framework. $C_B$ shows a well-defined minimum at small radii and then increases linearly with distance, which naturally reproduces the observed transition from Alfvénic to compressive fluctuations between $\sim$0.3–1 AU. The ratio $|\delta E_\parallel|/|\delta B_\perp|$ reveals opposite behaviors for the fast and slow modes: while the fast mode becomes more electrostatic with increasing distance, the slow mode evolves to a more magnetically dominated character. Expansion reduces the growth of their electromagnetic/compressive balance at large radii.  Our results demonstrate that solar wind expansion actively redistributes energy between magnetically compressive modes and purely transverse fluctuations with respect to the background magnetic field, playing a major role in shaping the radial evolution of wave dynamics throughout the inner heliosphere. 

\end{abstract}



\section{Introduction} 
\label{sec1}

First proposed by~\cite{Parker1958}, the solar wind has been a central subject of study for several decades, both through theoretical modeling and observational research. The interactions between plasma waves and turbulent fluctuations play a relevant role in shaping the large-scale dynamics of the solar wind~\citep{Bruno2013,Marino2023}. Over the past decades, multiple space missions have significantly advanced our understanding of the role of these multiscale phenomena in solar wind evolution across the heliosphere~\citep{Coleman1968,Marsch1982,Matteini2007}. More recently, the Parker Solar Probe~\citep{Fox2015} and Solar Orbiter~\citep{Mller2020} missions have extended these observations to the inner heliosphere, enabling detailed studies of the compressible energy cascade in solar wind turbulence~\citep{Brodiano2023}, proton temperature evolution~\citep{Rivera2024}, and other key processes. Despite these advances, the role of expansion in solar wind dynamics remains only partially understood.

An emblematic model incorporating solar wind expansion in the magnetohydrodynamic (MHD) equations is the Chew–Goldberger–Low (CGL) theory~\citep{CGL}. This model describes the thermodynamic evolution of a collisionless plasma without heat flux, in which the proton temperature anisotropy evolves inversely with the parallel plasma beta. As a result, CGL theory predicts an adiabatic cooling of the solar wind. However, \cite{Matteini2007} and \cite{Matteini2012} compared these predictions with proton temperature data from Helios observations between 0.3 and 1 AU, showing that the temperature anisotropy decreases more slowly with radial distance than predicted by CGL~\citep{Marsch1982,Hellinger2011}. Furthermore, this deviation distinguishes between fast and slow solar wind streams, with fast streams cooling more slowly than slow streams~\citep{Shi2022,Dakeyo2022}. This discrepancy indicates the presence of additional heating processes that act during the solar wind expansion to heat the plasma to observed temperatures, which are still not fully understood~\citep{Moya2012,Moya2014}. 

Since these additional heating processes are not captured by the standard CGL theory, several physical mechanisms have been proposed to play a role~\citep{Matthaeus2011,Bruno2013,Zank2018}. Among the relevant candidates, it has been proposed that Alfvénic waves can provide the necessary energy to heat fast currents, particularly for protons~\citep{Ofman2007} and minor heavy ions~\citep{Ofman2011, Moya2014}. During solar wind expansion, MHD waves, such as the Alfvén wave, play an important role in heating the solar wind, especially in the fast solar wind~\citep{Belcher1971b,Rville2020}. These wave modes have been extensively studied and characterized and correspond to the linear solutions of the ideal MHD equations for a homogeneous, infinitely extended plasma~\citep{Tu1995}. The three ideal MHD modes correspond to the Alfvén, fast magnetosonic, and slow magnetosonic modes. These modes exhibit distinct dispersive and compressive properties that determine their role in the transfer of energy across scales, such as magnetic compressibility $C_B$, the ratio of the parallel electric field to the perpendicular magnetic field $|E_\parallel|/|B_\perp|$, polarization $P$, among others~\citep{Gary_1986,Stix1992}. 

In the last decades, significant effort has been devoted to understanding the evolution of MHD waves in the solar wind and their contribution to solar wind heating. Early works studied their radial evolution in a spherically symmetric background magnetic field~\citep{Parker1965} and the propagation of Alfvén waves in a two-fluid model of the solar wind~\citep{Hollweg1973a}. Later,~\cite{Isenberg1984} extended these studies by investigating Alfvén wave propagation in a multi-ion plasma through a linear WKB approach. More recently,~\cite{Rivera2024} quantified the kinetic and thermal energy transfer from large-amplitude Alfvén waves to a fast solar wind stream, showing that MHD fluctuations play a key role in the fast solar wind heating and acceleration near the Sun. As MHD waves propagate and interact with the radially expanding solar wind over heliocentric distances, they play a major role in solar wind heating. Understanding how plasma expansion modifies their dispersion and propagation properties is essential for addressing the long-standing problem of solar wind heating dynamics.

To understand the role of expansion in plasma heating phenomena,~\cite{Velli1992} proposed the Expanding Box Model (EBM), a framework that reformulates MHD equations to account for the radial expansion of solar wind. The EBM introduces a non-trivial coordinate transformation to a non-inertial reference frame that co-moves with the plasma parcel at a constant velocity. In this frame, the spherical expansion of the solar wind is locally approximated by a Cartesian coordinate system, where the transverse coordinates to the solar wind's propagation direction are renormalized so that the plasma box maintains a constant volume. Consequently, plasma expansion effects no longer appear as explicit temporal dependencies but as additional non-inertial forces modifying the conservation of macroscopic plasma quantities in the MHD equations~\citep{Grappin1993,Grappin1996}. This formulation provides a convenient framework for numerical simulations and analytical studies. It limits possible memory constraints when simulating expanding plasma systems and provides a set of MHD equations in which expansion effects are expressed explicitly through additional terms. 

The EBM has since motivated extensive research into the role of expansion in plasma heating dynamics at both microscopic and macroscopic scales, both theoretically and in numerical simulations \citep{Liewer2001,Del_Zanna2015,Innocenti_2020,Echeverria-Veas2024}. 
More recently, the model has been generalized to describe solar wind acceleration near the Sun~\citep{Tenerani2017} and to extend its applicability to the quasi-linear regime of the solar wind to characterize microscopical scales~\citep{Seough2023}. Alfvén wave propagation within the EBM formalism has also been actively researched.~\cite{Nariyuki2015} studied nonlinear Alfvén waves in an accelerating EBM as solutions of the derivative nonlinear Schrödinger equation, while recently~\cite{Shi2020} studied Alfvénic fluctuations propagating in the solar wind with a fast-slow stream interaction through numerical simulations.

Despite these advances, the analytical characterization of the dispersive properties of linear MHD waves in an expanding solar wind has received comparatively little attention. In-situ observations have shown that the magnetic compressibility of solar wind fluctuations increases with heliocentric distance~\citep{Chen_2020,Zhao_2021}, suggesting a progressive enhancement of compressive fluctuations relative to Alfvénic ones as the plasma expands. Since magnetic compressibility is a key diagnostic for distinguishing different turbulent regimes in the solar wind~\citep{Chen_2020,Brodiano2023}, understanding its radial evolution is essential for understanding the transition from Alfvénic-dominated turbulence to the more compressive behavior farther from the sun. Nevertheless, the physical mechanisms underlying this behavior remain poorly understood. The effect of plasma expansion in the radial evolution of magnetic compressibility is of particular interest, as it directly modifies the propagation and energy distribution of the MHD modes that constitute MHD turbulence. In this context, the EBM provides an appropriate theoretical framework for investigating how expansion alters the dispersive properties of MHD waves, thereby allowing the study of whether the observed increase in magnetic compressibility can be explained by expansion acting on the underlying wave modes of the turbulent cascade.

The EBM offers significant analytical advantages by providing a set of MHD equations in which non-inertial forces represent the stretching of the plasma due to radial expansion~\citep{Grappin1993,Echeverria-Veas2023}, thereby providing a theoretical framework in which this problem can be addressed. Recently,~\cite{Saldivia_2025} studied the dispersion relation of MHD waves propagating in a radially aligned background magnetic field within the MHD-EBM framework. In this work, we extend this analysis to solar wind conditions by considering a Parker spiral structure of the background magnetic field, described by~\cite{Parker1958}. This approach allows us to characterize the propagation and dispersive properties of MHD waves in a solar-wind–like environment and to assess how expansion influences their radial evolution and observable signatures. Since properties such as wave polarization and magnetic compressibility have been extensively measured at various heliocentric distances, our results aim to provide a theoretical basis for interpreting their observed radial trends in the solar wind. In the following sections, we will linearize the MHD-EBM equations to obtain the corresponding dispersion relation in an expanding framework for a nonuniform background magnetic field. Using the deduced dispersion tensor, we will calculate the mode-dispersive properties and characterize their radial profiles. 

This paper will be organized as follows: in Section \ref{sec2}, we introduce the EBM formalism and the modifications to the ideal MHD set of equations (MHD-EBM). In Section \ref{sec3}, we linearize the MHD-EBM set of equations to obtain the wave equation for the ideal MHD normal modes of an expanding, solar wind-like plasma. In Section \ref{sec4}, we study the dispersive properties of the MHD-EBM modes. Finally, in Section \ref{sec5}, we summarize and conclude our main results.

\section{MHD and EBM}
\label{sec2}

In order to develop an analytical description of the effect of spherical solar wind expansion on the dispersion properties of MHD waves, we employ the EBM framework developed by~\cite{Velli1992} and~\cite{Grappin1993}. This formalism introduces a change of coordinates to a non-inertial, co-moving frame to describe the expansion of a plasma parcel from an initial position $R_0$, neglecting curvature effects through a Cartesian approximation. If the plasma expands away from a fixed, inertial system of reference $S$, the EBM introduces the transformation to a new system $S'$ co-moving with the plasma box at a constant velocity $V_0$, where the radial position of the plasma box at time $t$ is given by $R(t) = R_0 + V_0 t$. 
\begin{figure}
    \centering
    \includegraphics[width=0.8\linewidth]{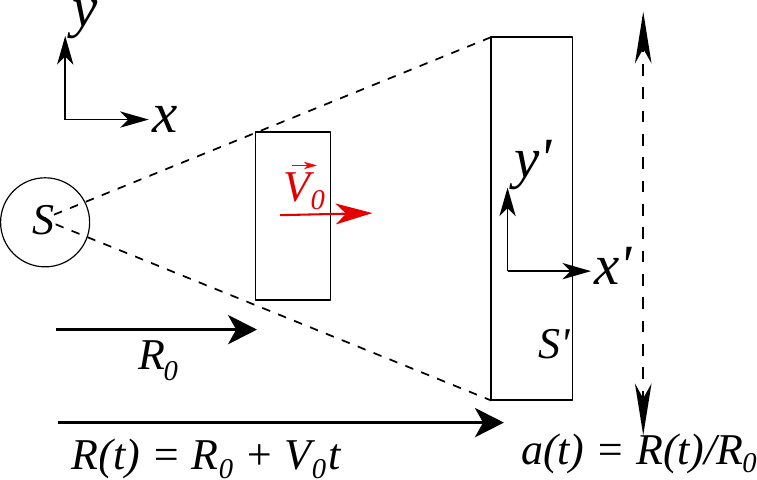}
    \caption{Coordinate transformation in the EBM. The radial expansion of the plasma is described through a Cartesian approximation. The plasma box expands away from a fixed reference system $S$ at a radial distance $R(t)$. The EBM introduces a non-inertial reference system $S'$, co-moving with the plasma box at a constant velocity $V_0$. The perpendicular coordinates are renormalized in the co-moving frame by the expansion parameter $a(t)$, maintaining a constant volume of the box.}
    \label{fig:1}
\end{figure}
We can write the coordinate transformation between systems $S$ and $S'$ through a Galilean transformation in the radial $(\hat x)$ axis and a renormalization in the transverse coordinates $\hat y$ and $\hat z$,
\begin{align}
    \frac{\partial}{\partial x} = \frac{\partial}{\partial x'}, \qquad \frac{\partial}{\partial y} = \frac{1}{a(t)} \frac{\partial}{\partial y'}, \qquad  \frac{\partial}{\partial z'} = \frac{1}{a(t)} \frac{\partial}{\partial z'},
\end{align}
where
\begin{equation}
    a(t) = \frac{R(t)}{R_0} = 1 + \frac{V_0}{R_0}t.
\end{equation} Here, $a(t)$ is a dimensionless parameter that maintains the constant volume of the plasma box through the contraction of the transverse coordinates. Figure \ref{fig:1} shows the relation between the reference systems $S$ and $S'$. 

The expansion parameter $a(t)$ allows us to study plasma expansion in a model fixed at a constant volume, as the expansion effects will be introduced when the spatial and temporal derivatives are written in the co-moving frame. Following~\cite{Grappin1993}, and ~\cite{Echeverria-Veas2023}, the relation of the spatial gradients and temporal derivatives between reference systems can be written as
\begin{align}
    \nabla & = \mathbb A^{-1}\cdot \nabla',\\ \qquad \nabla_v &= \mathbb A^{-1}\cdot \nabla_v',\\ \frac{\partial}{\partial t} &= \frac{\partial}{\partial t'} - \mathbf D \cdot \nabla',
\end{align}
where
\begin{align}
    \mathbb A = \begin{pmatrix}
        1 & 0 & 0\\
        0 & a(t) & 0\\
        0 & 0 & a(t)
    \end{pmatrix},&  & \mathbf D = V_0 \left (1 , \frac{y'}{R(t)},   \frac{z'}{R(t)}\right ).
\end{align}
In the co-moving frame, information about plasma expansion is stored in the $\mathbb A(t)$ tensor. As the perpendicular coordinates are renormalized (i.e., stretched and compressed), the expansion affects these coordinates, but no modifications are introduced to the radial axis, as expressed in the tensor components. Using the previous transformations, it is possible to write the ideal MHD equations in an expanding frame (\cite{Echeverria-Veas2023} and ~\cite{Grappin1993}, neglecting the primes in quantities)
\begin{equation}
    \frac{\partial n}{\partial t} + \nabla \cdot (n \mathbf u) = - \frac{2 \dot a }{a} n,
    \label{con}
\end{equation}
\begin{equation}
    \frac{\partial p}{\partial t} + \mathbf u \cdot \nabla p + \gamma p \nabla \cdot \mathbf u = -  \gamma \frac{2 \dot a}{a} p,
    \label{pre}
\end{equation}
\begin{equation}
    \begin{split}
            \frac{\partial \mathbf u}{\partial t} + (\mathbf u \cdot \nabla)\mathbf u + \frac{1}{8 \pi \rho} [(\mathbb A^{-1} \cdot \nabla )]B^2 - \\ \frac{1}{4 \pi \rho} [\mathbf B \cdot  (\mathbb A^{-1}\cdot \nabla)] \mathbf B + \frac{1}{\rho}\nabla p = - \frac{\dot a}{a} \mathbb T \cdot \mathbf u\,,
    \label{mom}
    \end{split}
\end{equation}
where
\begin{align}
\mathbb L = \begin{pmatrix}
    2 & 0 & 0\\
    0 & 1 & 0\\
    0 & 0 & 1
\end{pmatrix}, \qquad \mathbb T = \begin{pmatrix}
    0 & 0 & 0\\
    0 & 1 & 0\\
    0 & 0 & 1
\end{pmatrix}.
\end{align}
Note that we have assumed a scalar pressure equation for Equation \eqref{pre}, assuming a constant polytropic index $\gamma$. Here, the momentum equation describes the bulk flow velocity of the plasma. Given the small electron-to-proton mass ratio, this is approximated to the proton velocity. These equations are often coupled with Maxwell Equations, which have been developed in an expanding frame by~\cite{Echeverria-Veas2023}. Here, we will focus on the dispersion properties of linear MHD waves. We will only work with Faraday's Law in the EBM frame, written as
\begin{equation}
    (\mathbb A^{-1} \cdot \nabla )\times \mathbf E = -\frac{1}{c} \left ( \frac{\partial \mathbf B}{\partial t} + \frac{\dot a}{a} \mathbb L \cdot \mathbf B\right ),\label{fara}
\end{equation}
where we have neglected the Hall term. In the ideal MHD approximation, we have
\begin{equation}
     \frac{\partial \mathbf B}{\partial t} - [
            \nabla \times (\mathbf u \times \mathbf B)] = - \frac{\dot a}{a}\mathbb L \cdot \mathbf B.
            \label{bfl}
\end{equation}

We will use this system of expanding MHD equations to develop an analytical description of the radial evolution of the dispersion properties of MHD waves in the expanding solar wind. These equations have been mainly used in the literature to study expanding plasma phenomena, primarily through numerical simulations~\citep{Liewer2001,Grappin1996}.
Although solar wind expansion is inherently non-adiabatic, we use an adiabatic equation of state for pressure, as given by equation \eqref{pre}. This approximation is based on the fact that MHD waves occur on time scales that are short enough to neglect heat flow. This allows us to use an adiabatic equation of state for pressure. The term $\dot a/a$ on the right-hand side of Equations \eqref{mom}-\eqref{bfl} has usually been identified in literature as the inverse of the characteristic expansion frequency, where $\dot a/a = 1/\tau$, where $\tau$ is the expansion time scale~\citep{Nariyuki2015,Shi2020,Saldivia_2025}.

\section{Linear MHD-EBM waves in a non-uniform magnetic field}
\label{sec3}
\begin{figure*}[ht!]
    \centering
    \includegraphics[width=0.9\linewidth]{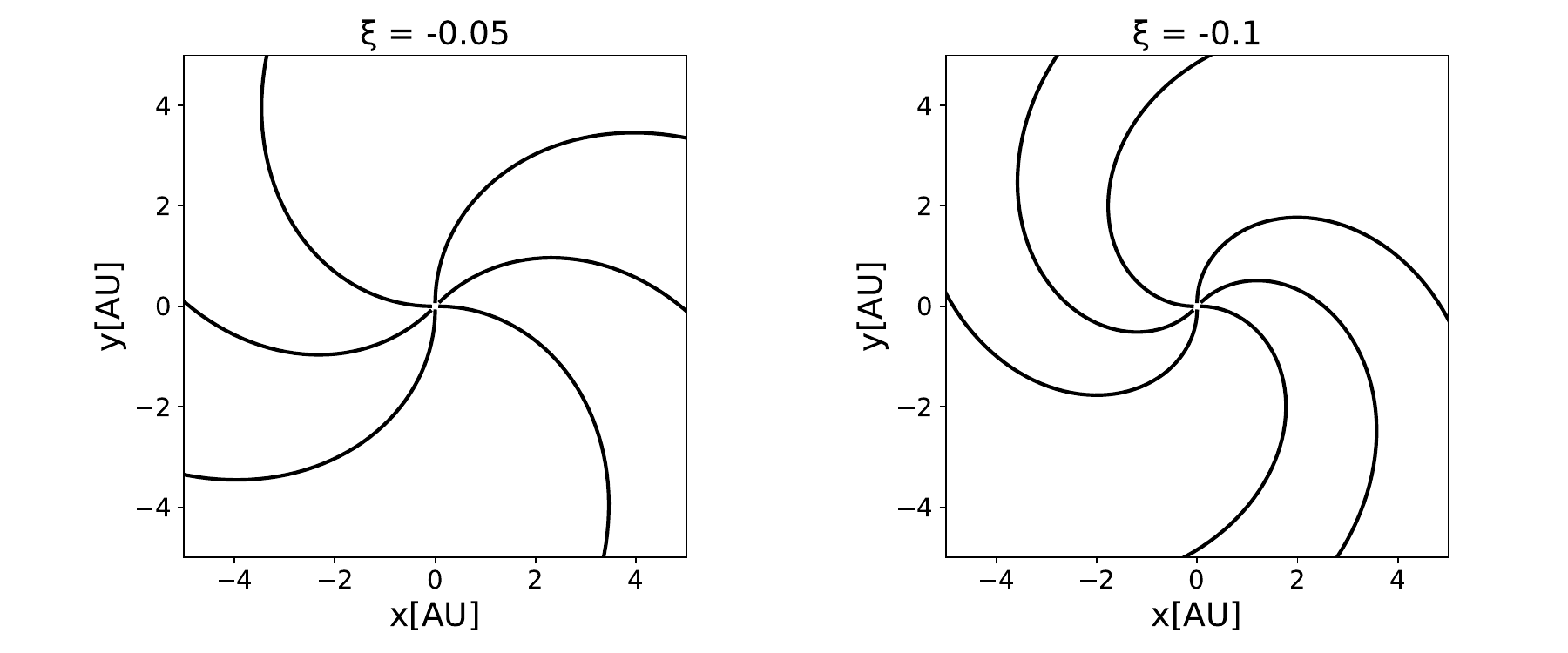}
    \caption{Background magnetic field lines represented in the x-y plane for each given value of the $\xi$ parameter, which represents fast solar wind (left) and slow solar wind (right). A larger magnitude of $\xi$ implies a more tightly spiraled magnetic field profile.}
    \label{fig:parkerspiral}
\end{figure*}
In this section, we derive the dispersion relations of ideal MHD waves in the radially expanding solar wind using Equations \eqref{con}, \eqref{pre},\eqref{mom}, and \eqref{bfl}. To accomplish this, we must choose a background magnetic field configuration to represent the spiral geometry of the interplanetary magnetic field.

Following the standard procedure for deriving linear plasma waves dispersion relations~\citep{Alfven1942}, we consider small-amplitude perturbations to the macroscopic plasma quantities of a perfectly conducting, infinitely extended plasma described by the ideal MHD-EBM equations. Each quantity, density $n$, pressure $p$, magnetic field $\mathbf B$ and velocity $\mathbf u$, is decomposed into a background quantity and a first-order small perturbation:
\begin{align}
    n & = n_{0} +  \delta n,   & p = p_{0} + \delta p,\\  \mathbf B &= \mathbf B_{0} + \delta  \mathbf B, & \mathbf u  = \mathbf u_{0} + \delta \mathbf u.
\end{align}
In this linear regime, we assume $\phi_{0} \gg \delta \phi$ for each quantity $\phi$, and model perturbations as traveling waves that can propagate in any direction, given by $\delta \phi(\mathbf r, t) = \delta \phi e^{i (\mathbf k\cdot \mathbf r - \omega t)}$. In the expanding frame, the background quantities are time- and space-dependent due to non-inertial forces that stretch and compress the plasma. Therefore, we first solve the zero-order differential equations to determine the background profiles, and then substitute these solutions into the first-order equations to obtain the wave equation whose roots will define the expanding dispersion relations $\omega(k,R)$. Assuming a background fluid velocity $\mathbf u_{0} = 0$, the zero-order differential equations can be solved analytically to obtain the background profiles
\begin{equation}
    n_{0} = \frac{\bar n_0}{a^2},\label{eq11}\\
\end{equation}
\begin{equation}
    p_{0} = \frac{\bar p_0}{a^{2 \gamma}},\label{eq12}\\
\end{equation}
\begin{equation}
    \mathbf B_{0} = \mathbb Z^{-1} \cdot \bar{\mathbf B_0},\label{eq13}\\
\end{equation}
where
\begin{equation}
    \mathbb Z = \begin{pmatrix}
        a(t)^2 & 0 & 0\\
        0 & a(t) & 0\\
        0 & 0 & a(t)
    \end{pmatrix},
\end{equation}
and $\mathbf B_0$ is the  background magnetic field, 
%revisar superindices indices
This is consistent with the well-known radial profiles of plasma quantities in the EBM frame~\citep{Del_Zanna2015,Innocenti_2019}. Note that the radial profile of the background magnetic field naturally resembles a Parker spiral structure of the heliosphere, modeled by~\cite{Parker1958} in spherical coordinates as
\begin{equation}
    B_r(R, \theta, \phi) = B_{0}(\theta, \phi_0) \left ( \frac{R_0}{R}\right )^2,\label{parkerr}
\end{equation}
\begin{equation}
    B_\theta(R,\theta,\phi) = 0,\label{parkert}
\end{equation}
\begin{equation}
    B_\phi(R,\theta,\phi) = B_{0}(\theta, \phi_0) \frac{\Omega}{v_{sw}} (R - R_0) \left (\frac{R_0}{R} \right )^2 \sin \theta,\label{parkerp}
\end{equation}
where $\Omega = 2.87 \cdot 10^{-6}$s${}^{-1}$ is the solar rotation frequency at the equatorial plane, $v_{sw}$ is the solar wind speed and $B_{0}(\theta, \phi_0)$ represents the background magnetic field at the starting position $R_0$. In the equatorial plane, the ratio between the azimuthal and radial magnetic field components is given by 
\begin{equation}
    \frac{B_\phi}{B_r} = - \frac{\Omega R(t)}{v_{sw}}.\label{ratiop}
\end{equation}
On the other hand, for the magnetic field profile given by the EBM formalism, following~\cite{Echeverria-Veas2024}, we can rewrite Equation \eqref{eq13} as
\begin{equation}
    \mathbf B_0 =\frac{B_{0x}}{a^2} (1, a \xi, a \xi),\label{field}
\end{equation}
where $\xi$ represents the ratio between the background magnetic field components,
\begin{equation}
    \xi = \frac{B_{0y}}{B_{0x}} = \frac{B_{0z}}{B_{0x}}.\label{xi}
\end{equation}

Comparing Equations \eqref{parkerr}-\eqref{parkerp} and \eqref{field} shows clearly that the two representations are analogous to each other. Hence, in Equation \eqref{field}, the dimensionless parameter $\xi$ quantifies the spiraling of the background magnetic field lines. In the $\xi \to 0$ limit, the field is purely radial, corresponding to a magnetic field configuration aligned with the solar wind flow. As $\xi$ increases in magnitude, the field becomes more azimuthal, describing the spiraling of the interplanetary magnetic field (IMF) induced by solar rotation. Identifying the ratio given by Equation \eqref{ratiop} at the reference position $R_0$ with the definition of $\xi$ given by Equation \eqref{xi}, we obtain
\begin{equation}
    \xi =-\frac{\Omega R_0}{v_{sw}}.    
\end{equation}
Thus, $\xi$ measures the local pitch angle of the field at the starting position $R_0$, as the implicit radial evolution $R(t)$ is expressed in the $a(t)$ term. We consider two cases of a plasma parcel expanding from $R_0 = 0.1$ AU with solar wind velocities $v_{sw} = 400$km/s and $v_{sw} = 750$km/s, representing slow and fast solar wind, respectively. Corresponding $\xi$ values in each case are $\xi = -0.1$ and $\xi = - 0.05$. Figure \ref{fig:parkerspiral} shows a 2D representation of this background magnetic field profile, for the two given values of $\xi$.
TThis description allows us to characterize the effect of radial plasma expansion on the radial evolution of dispersive properties of MHD waves in the solar wind, adopting a background magnetic field representative of actual solar wind conditions.

\section{Results: Wave properties in the expanding Parker spiral}
\label{sec4}

\subsection{Dispersion Relation and Eigenfrequencies}

\begin{figure*}[ht!]
    \centering
    \includegraphics[width=0.7\linewidth]{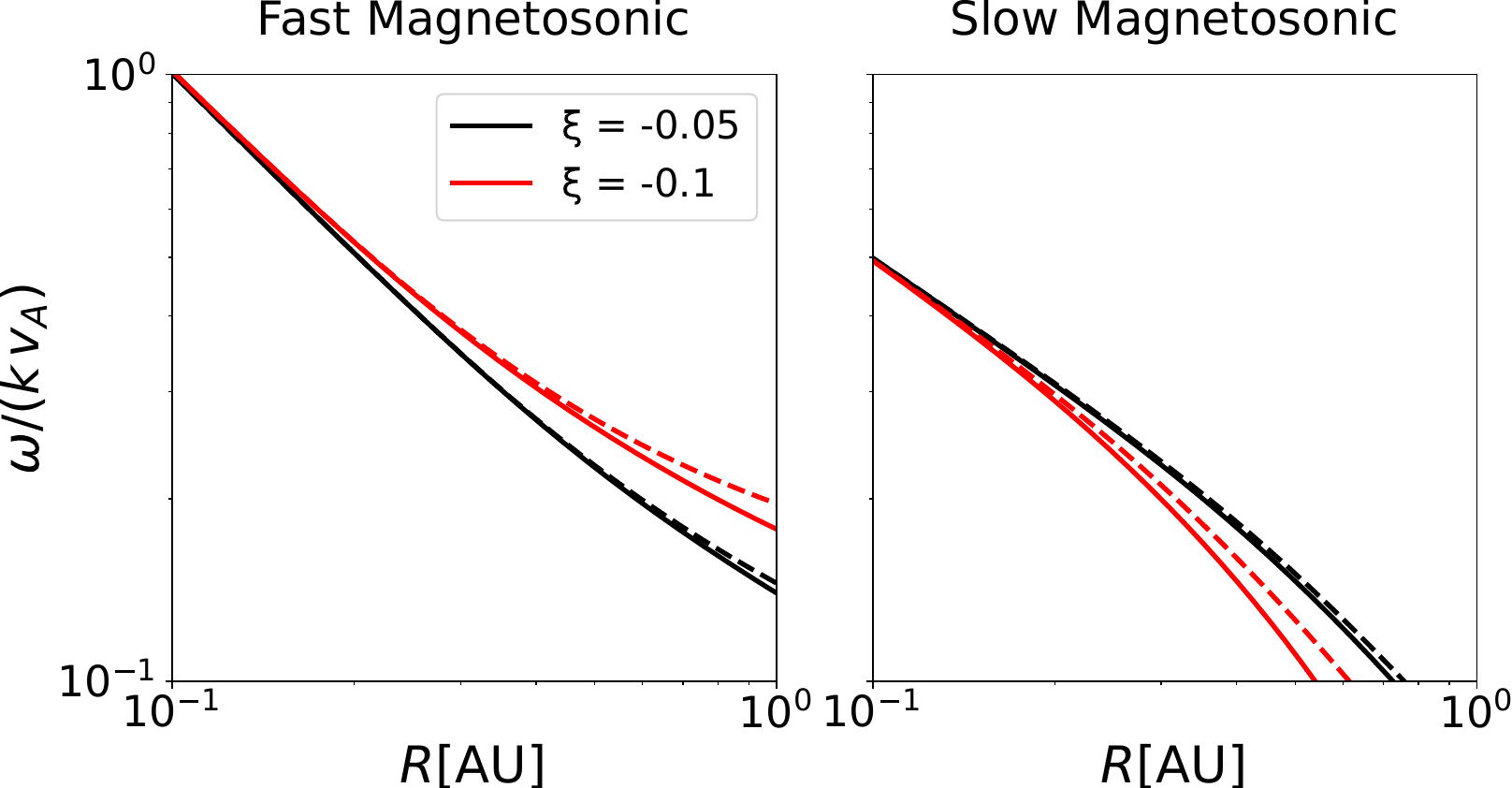}
    \caption{Normalized phase speeds of fast magnetosonic, slow magnetosonic waves as a function of heliocentric distance $R$, computed for a Parker-spiral background magnetic field and a wave vector aligned with the x-axis. The parameter $\xi$ distinguishes the slow solar wind case ($\xi = -0.1$, red line) from the fast solar wind case ($\xi = -0.05$, black line). Solid lines show the phase speeds obtained within the MHD-EBM framework, including plasma expansion from first principles as given by Equations \eqref{fast}-\eqref{alfven}, while dashed lines show the corresponding values derived from a parametric rescaling of the non-expanding case, as given by Equations \eqref{disp1}-\eqref{disp3}. Here $R_0 = 0.1$ AU.}
    \label{fig:disp}
\end{figure*}

In this section, we focus on the dispersion properties of linear MHD waves expanding in a magnetic field with the configuration given by Equation \eqref{field}. To do so, we must first write the dispersion relation in matrix form for this system, as information about wave properties can be derived from the corresponding dispersion tensor. Using Equation \eqref{field} and assuming a radial wave propagation ($\mathbf k = k \hat x$), inserting Equations \eqref{eq11}-\eqref{eq13} into the first-order MHD-EBM equations yields the dispersion relation $\mathbb D \cdot \delta \mathbf u = 0$. This is solved analytically in a weak expansion limit, where a Fourier transform is performed by treating the scale factor $a(t)$ as quasi-constant, since the characteristic expansion frequency $1/\tau$ is many orders of magnitude smaller than MHD wave frequencies. Thus, solar wind expansion evolves slowly compared with wave oscillations (see~\cite{Saldivia_2025} for a detailed derivation). The resulting components of the dispersion matrix $\mathbb D$ are 
\begin{align}
    D_{xx}& = \omega^2/k^2 -  \tilde v_A^2\frac{ \xi^2R}{R_0} \left (1 + \frac{R}{R_0} \right) - \tilde v_S^2 ,\\
    D_{xy}& =D_{xz} =      D_{yx}  =  D_{zx} =  \frac{\tilde v_A^2 \xi R}{R_0}  ,\\
    D_{yy} & = D_{zz} = \omega^2/k^2 - \tilde v_A^2,\\
    D_{yz} & = D_{zy} = 0.
\end{align}
Here, the Alfvén and sound speeds have been rewritten to implicitly include their radial decay due to plasma expansion. Thus, the main effect of radial expansion on MHD waves is a radial decay of plasma characteristic speeds, expressed as
\begin{equation}
    \tilde v_A(R)^2 = \frac{B_{0x}^2}{4 \pi \bar n_o m a^2} = v_A^2\frac{R_0^2}{R^2},
\end{equation}
\begin{equation}
    \tilde v_S(R)^2 = \frac{\gamma \bar p_0}{\bar n_0 m} a^{2 - 2 \gamma} =v_S^2  \frac{R_0^{2\gamma - 2}}{R^{2\gamma  - 2}},
\end{equation}
where $v_A$ and $v_S$ correspond to the Alfvén and sound speeds in the non-expanding limit, or at $t = 0$. Note that, in this framework, the Alfvén speed depends only on the radial component of the background magnetic field. This is because the modifications due to the spiral geometry of the background magnetic field are associated with the $\xi$ terms that appear in the components of the dispersion matrix. The dispersion relation of this system has non-trivial solutions when $\det(\mathbb D) = 0$, whose roots will yield the corresponding eigenfrequencies of the ideal MHD-EBM system as
\begin{equation}
    \begin{split}
        \omega^2 & = \frac{k^2}{2} (\tilde v_A^2(1 + a \xi^2 (1 + a)) + \tilde v_S^2   + [4a  \xi^2 (a -1) \tilde v_A^4 \\ & + (\tilde v_S^2 + \tilde v_A^2 (1 + a \xi^2 (1+a)))^2 - 4 \tilde v_A^2 \tilde v_S^2]^{1/2}),\label{fast}        
    \end{split}
\end{equation}
\begin{equation}
    \begin{split}
        \omega^2 & = \frac{k^2}{2} (\tilde v_A^2(1 + a \xi^2 (1 + a)) + \tilde v_S^2    - [4a  \xi^2 (a -1) \tilde v_A^4 \\ & + (\tilde v_S^2 + \tilde v_A^2 (1 + a \xi^2 (1+a)))^2 - 4 \tilde v_A^2 \tilde v_S^2]^{1/2}),\label{slow}        
    \end{split}
\end{equation}\begin{equation}
    \omega^2 = k^2 \tilde v_A^2. \label{alfven}
\end{equation}
Here, Equations \eqref{fast},\eqref{slow} and \eqref{alfven} correspond to the three eigenmodes of the expanding ideal MHD dispersion relation, representing the fast magnetosonic, slow magnetosonic and Alfvén modes, respectively, under a non-uniform background magnetic field resembling the Parker spiral structure of the solar wind. As wave propagation is assumed parallel to the radial axis, Equation \eqref{alfven} shows the typical dispersion relation of a shear-Alfvén wave, where the Alfvén speed decays as $\tilde v_A(R) \propto 1/R$. Since the Alfvén mode is purely transverse and does not depend on the field geometry given by the $\xi$ parameter, and since this mode is not compressive, we will now focus on the dispersive properties of the fast and slow magnetosonic modes. These depend explicitly on the expanding parameter $a(t)$ and on $\xi$. In these modes, both the shape of the background magnetic field and the coordinate stretching inherent to the EBM framework modify the analytical form of the eigenfrequencies through the terms coupled to the Alfvén speed. These expressions differ from those obtained in the case of a purely radial background magnetic field, where the same form as the non-expanding form is recovered by introducing only a radial dependence of plasma speeds~\citep{Saldivia_2025}. For $\xi = 0$, we recover this radial case for $\mathbf k = k \hat x$.

\begin{figure*}[ht!]
    \centering
    \includegraphics[width=0.7\linewidth]{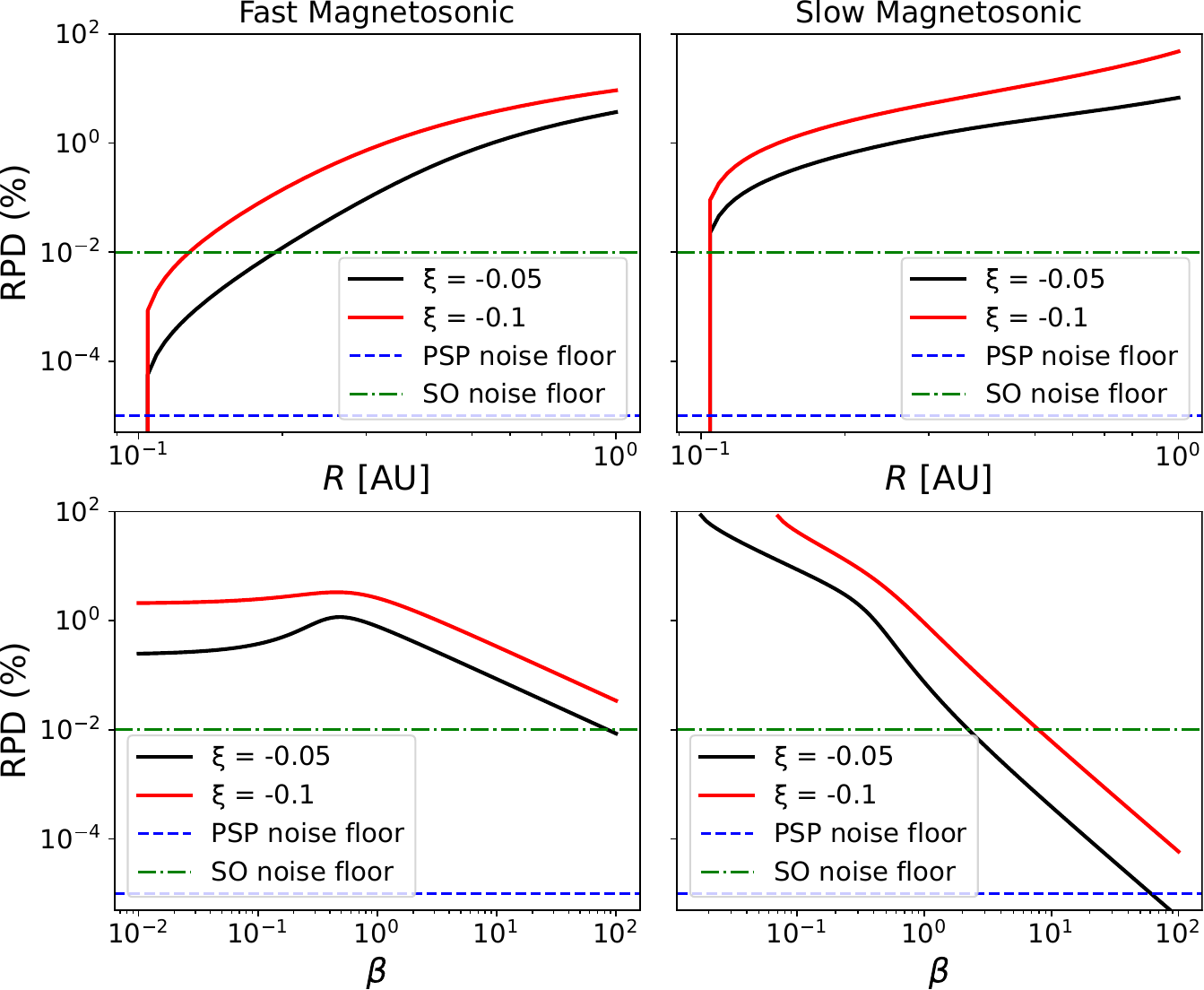}
    \caption{Relative percentage difference between expanding and non-expanding phase speeds for fast and slow modes, shown vs heliocentric distance $(\beta = 0.25)$ and plasma beta ($R = 0.5$ AU), for both $\xi$ values. Horizontal lines indicate noise floor percentage assuming $B_0 \sim 150$nT at $R_0 = 0.1$ AU.}
    \label{fig:rpd}
\end{figure*}

In Appendix \ref{appendixa}, we present a direct derivation of the dispersion relation in a non-expanding frame using the standard MHD equations. From Equations \eqref{fast}-\eqref{alfven}, the non-expanding eigenfrequencies are recovered in the limit $a = 1$. In the literature, radial profiles of plasma quantities, such as the Alfvén speed or the background magnetic field, are often assumed to capture solar wind expansion in a simplified manner. Thus, Equations \eqref{disp1}-\eqref{disp2} represent the application of such radial dependencies to the non-expanding dispersion relations, providing a direct comparison between the first-principles derivation within the EBM framework and the plasma quantities rescaling approach. While the Alfvén mode is equivalent in both cases, indicating that it is governed by the local Alfvén speed decay, the magnetosonic modes differ in their analytical expressions. This discrepancy arises because the EBM formalism modifies not only the radial profiles of the background quantities, but also the structure of the spatial and temporal operators, introducing anisotropic fluid stretching into the MHD equations~\citep{Echeverria-Veas2023}. This modifies the magnetic tension term and couples transverse components of the perturbations, which cannot be reproduced by a parametric rescaling of plasma quantities.

Figure \ref{fig:disp} shows the radial evolution of the normalized phase velocity, $\omega/(k v_A)$, for the fast and slow magnetosonic modes in logarithmic scale, with $v_A/v_S = 2$, which implies a plasma beta of $\beta = 0.25$, and $R_0 = 0.1$ AU. We consider two values of $\xi$ (-0.05 and -0.1), representing fast and slow solar wind conditions, respectively.  Solid lines show the expanding phase speeds obtained from the MHD-EBM framework as given by Equations \eqref{fast}-\eqref{alfven}, while dashed lines represent the phase velocity as obtained from the parametric rescaling of plasma quantities in the non-expanding case, as given by Equations \eqref{disp1}-\eqref{disp3}. As the solar wind expands with heliocentric distance $R$, the phase velocity (and therefore the wave frequency) of the fast and slow magnetosonic modes decreases, as Equations \eqref{fast}-\eqref{alfven} suggested. Figure \ref{fig:rpd} shows the relative percentage difference (RPD) between the expanding and non-expanding cases for the fast and slow magnetosonic modes, plotted as a function of heliocentric distance for $\beta = 0.25$ (top) and as a function of plasma beta for $R = 0.5$ AU (bottom), for each $\xi$ value. Additionally, we compare this RPD with the percentage associated with the instrumental background noise level of the FIELDS magnetometer onboard the Parker Solar Probe (PSP) mission, which is equipped with a  with a noise floor of 0.01 pT Hz$^{-1/2}$~\citep{Larosa_2022}, and the MAG instrument of the Solar Orbiter (SolO) spacecraft, with a noise floor of 10 pT~\citep{Horbury2020}. We assume a mean background magnetic field of $B_0 \sim 150$nT at $R_0 = 0.1$ AU. As these relative percentage differences exceed the noise level at large distances and small betas, we conclude that the phase-velocity modification associated with plasma expansion on these modes is significant enough to be measured.

Several other plasma diagnostics can be computed from the dispersion relation, such as wave polarization, magnetic compressibility, or the electric-to-magnetic amplitude ratio~\citep{Moya_2022,Villarroel-Sepulveda2023}. Following ~\cite{Stix1992}, polarization in an orthogonal basis to the background magnetic field is $P = 0$ by construction in this case, implying that MHD waves conserve their linear polarization in the EBM frame and that this polarization is unaffected by expansion. 

\begin{figure*}[ht!]
    \centering
    \includegraphics[width=0.8\linewidth]{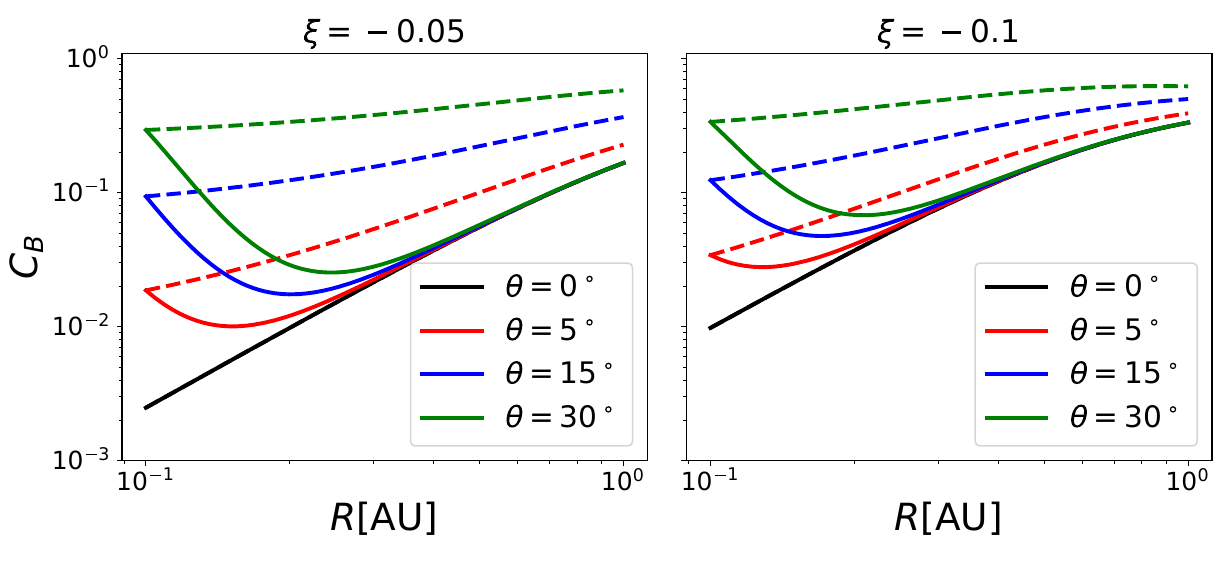}
    \caption{Magnetic compressibility $C_B$ as a function of heliocentric distance $R$, computed for a Parker-spiral background magnetic field where a wave vector $\mathbf k$ is oriented at an angle $\theta$ with respect to the background magnetic field, and $R_0 = 0.1 $AU. The value of $\xi$ distinguishes between fast solar wind $(\xi = - 0.05$) and slow solar wind $(\xi = - 0.1$). Solid lines represent the expanding case as obtained from the MHD-EBM framework given by Equation \eqref{cb}, while dashed lines represent the non-expanding case given by Equation \eqref{cbnoexp}.}
    \label{fig:compressibility}
\end{figure*}

\subsection{Magnetic Compressibility}
Following~\cite{Gary_1986}, we compute the magnetic compressibility of each mode as a function of the magnetic field perturbations derived in Appendix \ref{appendixc} and given by Equations \eqref{eqbx}-\eqref{eqbz}. This quantity measures the degree to which magnetic fluctuations are compressive. It is defined as
\begin{equation}
    C_B =\frac{|\delta B_\parallel|}{|\delta B|}, \label{cbd}
\end{equation}
where the parallel component of the magnetic field perturbation $\delta B_\parallel$ can be written as $\delta B_\parallel = \delta \mathbf B \cdot {\mathbf B_0}/B_0$. For the slow and fast magnetosonic modes, we can insert the expanding Faraday's law components given by Equations \eqref{eqbx} and \eqref{eqbz} into Equation \eqref{cbd}, obtaining
\begin{equation}
    C_B = \frac{\displaystyle {\sin^2 \theta} + a^8 \xi^2 {\cos^2 \theta} - {2 a^4 \xi\sin \theta \cos \theta }}{\displaystyle {\sin^2 \theta} + a^6 {\cos^2 \theta}  + 2 \xi^2 a^2 {\sin^2 \theta} + 2 \xi^2 a^8 \cos^2 \theta }.\label{cb}
\end{equation}
We also evaluate the magnetic compressibility for the non-expanding MHD system, assuming a radial profile of the background magnetic field amplitude as described by Equation \eqref{field}. Using Equation \eqref{cbd} and the non-expanding Faraday's law, we obtain 
\begin{equation}
    C_B^{(0)} = \frac{\sin^2 \theta + a^2 \xi^2 \cos^2 \theta - 2 a \xi \cos \theta \sin \theta}{1 + 2 \xi^2 a^2}.\label{cbnoexp}
\end{equation}
Figure \ref{fig:compressibility} shows the radial evolution of the magnetic compressibility of magnetosonic waves for a representative set of parameters. We choose $R_0 = 0.1$ AU as the reference distance, so that the radial evolution of compressibility is limited to the inner heliospheric region, where in situ measurements allow us to directly compare MHD-EBM predictions with current observations. We chose the values of $\xi = -0.1$ and $\xi = -0.05$ AU to represent different degrees of magnetic field spiraling associated with slow and fast solar wind streams, respectively, where the associated speeds of the solar wind streams ($v_S = 450$ and $750$ km/h) lie within the observed range~\citep{Bruno2013}. In Figure \ref{fig:compressibility}, solid lines correspond to the expanding case described by Equation \eqref{cb}, while dashed lines represent the non-expanding case described by Equation \eqref{cbnoexp}, both in logarithmic scale for different propagation angles $\theta$. For $\theta = 0^\circ$ and $5^\circ$, $C_B$ increases by roughly one order of magnitude, from $\sim 10^{-2}$ to $\sim 10^{-1}$, as the solar wind expands, for both fast and slow wind conditions. This is consistent with in-situ observations across heliocentric distances~\citep{Chen_2020, Zhao_2021}.

At larger propagation angles, the two models differ substantially in their radial evolution. While the non-expanding case exhibits an approximately power-law radial dependence, the MHD-EBM formulation predicts the appearance of a minimum in $C_B$ at a certain heliocentric distance. The position of this minimum can be obtained analytically from Equation \eqref{cb} as 
\begin{equation}
    R_{\text{min}} = R_0\left(\frac{ \tan  \theta }{\xi}\right)^{1/4}.
\end{equation}
The location of this minimum in space depends explicitly on the expansion geometry and propagation angle. It indicates a radial region during solar wind expansion, the size of which depends on $R_0$, where the contribution of compressive magnetic fluctuations temporarily decreases. This could leave an observable signature in in situ measurements of magnetic compressibility at the inner heliosphere. 

For the parameters considered here, beyond $R \sim 0.3$–$0.4$ AU up to $\sim$1 AU, the magnetic compressibility $C_B$ resumes a monotonic increase that is well described by a power-law dependence. This increase is consistent with in-situ measurements indicating a transition from a predominantly Alfvénic, magnetically dominated turbulent regime to a more hydrodynamic-like state, characterized by enhanced compressibility and reduced magnetic anisotropy~\citep{Bruno2013,Telloni2022,Chen_2020}. Since magnetic compressibility reflects the relative energy carried by the MHD wave modes in the turbulent cascade~\citep{Chen2016}, this result suggests that the observed transition may be partly driven by solar wind expansion.

Based on PSP measurements in the inner heliosphere, although with significant dispersion,~\cite{Chen_2020} reported a dependence of $C_B \propto R^{1.68 \pm 0.23}$ from observational data of the solar wind. To quantify the radial evolution predicted by this model, we fitted power-law scalings directly to the theoretical profiles shown in Figure \ref{fig:compressibility} using linear least-squares regression in logarithmic space. The fits were performed over the radial range $R \geq 0.3$ AU, where the solutions exhibit a linear trend in logarithmic space, compatible with a power-law behavior $C_B \propto R^\alpha$. A power-law exponents were obtained from a linear regression in $(\log R, \log C_B)$ space. For all cases considered, fittings are associated with a coefficient of determination of $\text R^2 \sim 0.99$.

For the non-expanding case, this procedure yields a radial profile of $C_B^{(0)} \propto R^{1.21}$ for $\theta = 5^\circ$. This angle represents quasi-parallel fluctuations, which are most relevant to the commonly observed solar wind conditions~\citep{Bruno2013}. As $\theta$ increases, the fitted slope systematically decreases and the extent of the log-log linear regime is reduced, obtaining $C_B^{(0)} \propto R^{0.72}$ for $\theta = 15^\circ$ and $C_B^{(0)} \propto R^{0.36}$ for $\theta = 30^\circ$, performing the fit from 0.3 AU to 1 AU.

\begin{figure*}[ht!]
    \centering
    \includegraphics[width=0.8\linewidth]{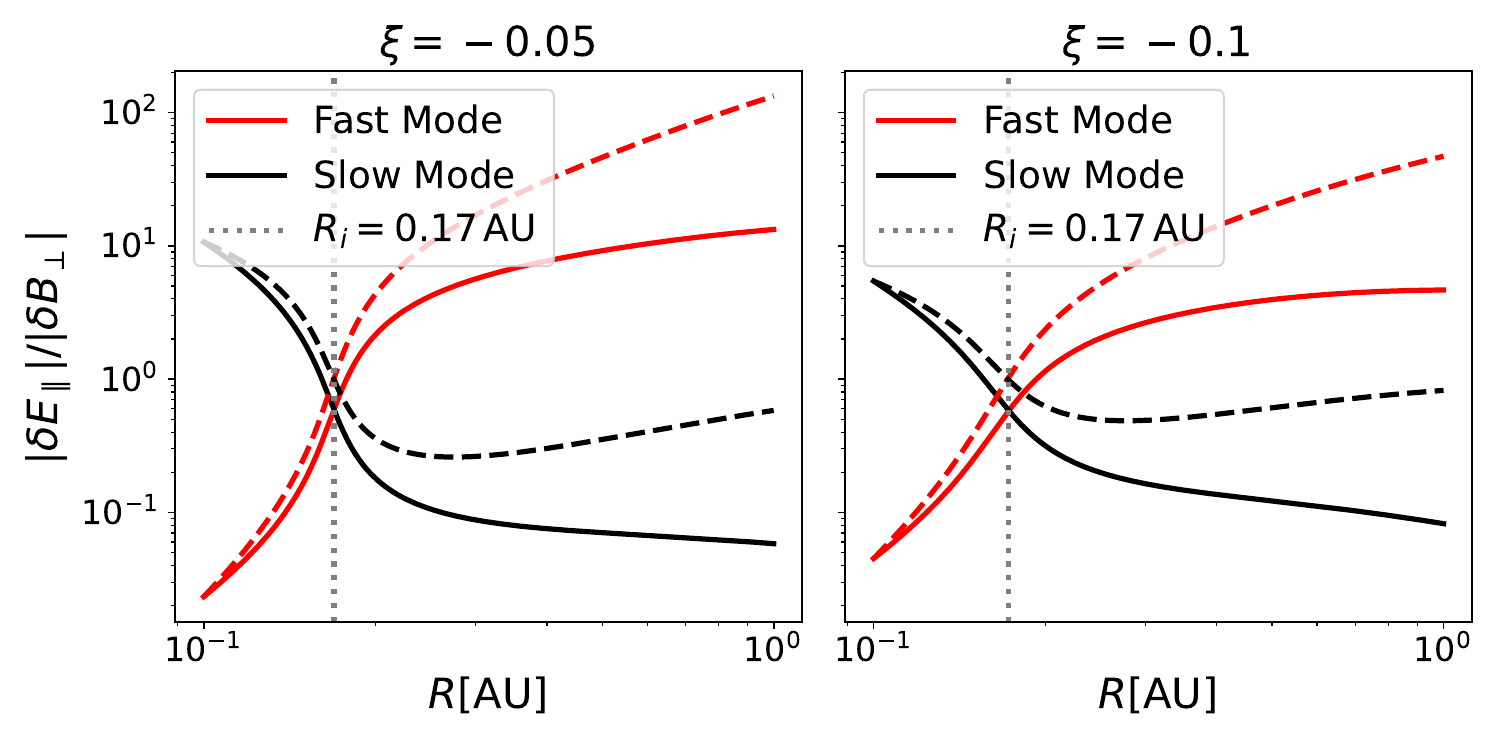}
    \caption{Parallel electric field to perpendicular magnetic field ratio $|\delta E_\parallel|/|\delta B_\perp|$ as a function of heliocentric distance $R$, computed for a Parker-spiral background magnetic field and a wave vector aligned with the x-axis. Here, $R_0 = 0.1$ AU. Solid lines represent the expanding case as obtained from the MHD-EBM framework given by Equation  \eqref{ratio}, while dashed lines represent the non-expanding case from standard MHD assuming the radial decay of plasma quantities, for both slow solar wind $(\xi = - 0.1$, red line) and fast solar wind $(\xi = - 0.1$, red line). The dotted line represents the intersection point between the curves, occurring at $R_i \sim 0.17$ AU in both cases for this set of parameters.}
    \label{fig:ratio}
\end{figure*}

When plasma expansion is included, the fitted exponents increase significantly. In particular, for fast solar wind, we obtain $C_B \propto R^{1.66}$ at $\theta = 5^\circ$, which is in close agreement with the observational trend of $C_B \propto R^{1.68 \pm 0.23}$ reported by~\cite{Chen_2020}. This suggests that plasma expansion plays a major role in the radial evolution of magnetic compressibility in the solar wind. For slow solar wind at $\theta = 5^\circ$ the scaling is $C_B \propto R^{1.35}$.  
For larger $\theta$ angles, the fitted slope decreases slightly as the linear range shifts toward larger heliocentric distances, yielding $C_B \propto R^{1.63}$, $C_B \propto R^{1.55}$ for fast solar wind at $\theta = 15^\circ$ and 30$^\circ$, respectively, and $C_B \propto R^{1.15}$ and $C_B \propto R^{1.11}$ for slow solar wind. In all cases, the fit is performed over the radial interval from 0.3 AU to 1 AU. Although observational data scatter makes it difficult to confirm the existence of the minimum predicted by the expanding solutions, the overall radial trends derived within the MHD–EBM framework provide a quantitatively consistent description of heliospheric observations, particularly at heliocentric distances beyond 0.3 AU~\citep{Tu1995,Bruno2013}.

\subsection{Electric-to-magnetic amplitude ratio}

Another quantity we can compute is the parallel electric field amplitude to perpendicular magnetic field amplitude ratio, ${|\delta E_\parallel|}/{|\delta B_\perp|} $, which indicates the field-polarization properties of each mode~\citep{Gary1993}. This can be written as
\begin{equation}
    \begin{split}
        \frac{|\delta E_\parallel|}{|\delta B_\perp|} = \frac{\omega}{kc} \frac{|\delta E_x + a \xi (\delta E_y + \delta E_z)|}{\displaystyle{\sqrt{K(a,\xi)}}}.\label{ratio}
    \end{split}
\end{equation}
Here, $K$ is given by Equation \eqref{K}, and a detailed derivation of this quantity can be found in Appendix \ref{appendixd}. In Equation \eqref{ratio}, the components of the perturbed electric field $\delta \mathbf E$ can be obtained as the cofactors of the dispersion tensor $\mathbb{D}$ associated with the linearized MHD-EBM system. As $\mathbb D \cdot \delta \mathbf E = 0$, each component can be written as $\delta E_i = C_i(\omega, k, R, \xi),$
where $C_i$ is the cofactor of the $i$-th column of $\mathbb{D}$. Thus, the polarization of the electric field for each eigenmode is directly obtained from $\mathbb{D}$, without assuming any particular propagation geometry. The result given by Equation \eqref{ratio} depends on the phase velocity $\omega/k$ associated with each mode, and can therefore be evaluated for each MHD-EBM mode. Since pure Alfvén waves are transverse to the background magnetic field, their parallel electric field component $\delta \mathbf E_\parallel$ vanishes, yielding $|\delta E_\parallel|/|\delta B_\perp| = 0$ for the Alfvén mode. 

Figure~\ref{fig:ratio} shows the radial evolution of the $|\delta E_\parallel|/|\delta B_\perp|$ ratio for the fast and slow magnetosonic modes, in logarithmic scale. The solid lines correspond to the expanding case described by Equation~\eqref{ratio}, while the dashed lines represent the non-expanding case in which the background magnetic field is given by Equation~\eqref{field}. For both the non-expanding and expanding cases, for each $\xi$ value, the ratio of the fast mode increases with heliocentric distance. It evolves from low electromagnetic character near the Sun with $|\delta E_\parallel|/|\delta B_\perp| \gg 1$ beyond $\sim 0.3$ AU, signaling a transition toward a regime where the electric response becomes larger than the magnetic one, consistent with Faraday's Law. Meanwhile, the ratio of the slow mode decreases with distance, indicating a transition to a more magnetically dominated regime. These opposite trends reflect the different radial scalings of the background Alfvén and sound speeds, which scale as $v_A \propto R^{-1}$ and $v_S \propto R^{1-\gamma}$ in the expanding background~\citep{Saldivia_2025}. These scalings modify the phase relations between the field components of each mode. Consequently, a transition emerges where the two ratios become equal at a certain heliocentric distance, such that  $|\delta E_\parallel (R_i)|/|\delta B_\perp (R_i)|_{\rm fast} = |\delta E_\parallel(R_i)|/|\delta B_\perp (R_i)|_{\rm slow}$. For the reference distance $R_0 = 0.1$ AU and the values of $\xi$ considered before, this condition yields 
$R_i \sim 0.17$AU for the set of parameters considered here, depending on the choice of $R_0$. Beyond this radius, the fast mode ratio overtakes the slow mode ration, with $|\delta E_\parallel|/|\delta B_\perp|_{\rm fast} > |\delta E_\parallel|/|\delta B_\perp|_{\rm slow}$ diverging by two orders of magnitude at $R = 1$ AU for both fast and slow wind conditions.

Beyond $\sim 0.3$–$0.4$ AU, the radial evolution of the amplitude ratios in the expanding case departs significantly from the non-expanding case, by about one order of magnitude in each case. While the non-expanding case shows a continued growth of $|\delta E_\parallel|/|\delta B_\perp|$ with radial distance for both the fast and slow modes, the inclusion of plasma expansion suppresses this increase, leading to a markedly slower radial evolution. This divergence indicates that solar wind expansion reduces the coupling imposed by Faraday’s law between electric and magnetic fluctuations by modifying the phase velocity and polarization of the modes. Although the present analysis focuses on magnetosonic modes, this expansion-driven reduction of the electric-to-magnetic coupling mirrors the observed radial decrease of Alfvénicity in solar wind fluctuations \citep{Bruno2007,DAmicis2021}, suggesting that expansion acts as a general regulator of electromagnetic correlations across different modes. In our case, expansion modifies the balance between $\delta \mathbf E_\parallel$ and $\delta \mathbf B_\perp$, leading to slower evolution of the amplitude ratio, with a decrease for the slow mode. Thus, the evolution predicted by the MHD-EBM framework shows expansion limiting the radial growth rate of $|\delta E_\parallel|/|\delta B_\perp|$, leading to a slower evolution of the electromagnetic character of compressive fluctuations.

\section{Discussion and Conclusions}
\label{sec5}

In this work, we have derived the dispersive properties of the three normal modes of ideal MHD in an expanding frame within the EBM framework, assuming a non-uniform background magnetic field consistent with the Parker spiral structure of the heliosphere. Starting from the MHD-EBM equations, we have constructed the dispersion tensor $\mathbb D$, from which we analytically obtained the eigenfrequencies $\omega(k, R)$, the magnetic compressibility $C_B$, and the ratio between the parallel electric field and the perpendicular magnetic field $|E_\parallel|/|B_\perp|$, following standard procedures in literature. This analytical framework allows us to characterize the radial evolution of these quantities in the expanding solar wind, thus quantifying how both the spiral topology and the radial plasma expansion modify wave propagation and dispersion properties across heliocentric distances, in that way, offering an analytical framework to study the evolution of dispersive properties of waves in the solar wind. Modeling the background magnetic field using the MHD-EBM allowed us to define a $\xi$ parameter that describes the spiraling of magnetic field lines. Thus, this analysis can distinguish between different solar wind streams. 

Our results show that the modification of the MHD-EBM eigenfrequencies $\omega(k, a, \xi)$ arises not only from the radial decay of the characteristic plasma speeds (Alfvén and sound speeds), which follow from the background profiles of plasma quantities, but also from additional terms arising from the transverse stretching inherent to the EBM frame. Thus, this deduction can capture new effects that do not appear in a rescaling of plasma speeds, in contrast to the radial background magnetic field case~\citep{Saldivia_2025}, as the EBM does not include non-inertial forces along the radial axis. The difference between the phase speeds obtained from the first-principles expanding framework and those from the standard non-expanding radial rescaling of quantities reaches over $\sim 5-10\%$ at $R \geq 0.5$ AU and for $\beta \leq 0.3$ for fast and slow modes, exceeding the noise floor levels of Parker Solar Probe and Solar Orbiter. This indicates that expansion-driven modifications on wave properties are large enough to be directly measurable with these missions. This theoretical framework could also lay the groundwork for studying scenarios in which a strongly anisotropic expansion must be taken into account. 

This difference is absent for the Alfvén mode eigenfrequency, which follows the well-known dispersion relation for shear Alfvén waves but exhibits parametric radial decay in Alfvén speed. This can be interpreted as a consequence of computing the eigenfrequencies for a wave vector aligned with the radial axis, $k = k \hat x$. Since shear Alfvén waves are transverse to the background magnetic field, this approximation does not capture additional terms depending on the spiral structure. However, due to the analytical difficulty of this deduction in general oblique wave propagation, this limitation could be overcome through numerical solutions. 

Within this framework, the MHD modes exhibit the expected vanishing of their frequencies with increasing heliocentric distance. In this expanding plasma, the background geometry and gradients evolve with the field amplitudes, altering the propagation operator itself rather than merely the equilibrium values. Consequently, the Alfvén speed appearing in the dispersion relations depends exclusively on the decay of the radial component due to magnetic flux conservation in an expanding radial flux tube. The transverse components, decaying more slowly, do not enter symmetrically in the restoring magnetic tension and instead modulate the anisotropic coupling terms in the dispersion matrix, yielding additional $a(t)$ terms, such as the off-diagonal components—that cannot be captured by applying radial scalings into the non-expanding dispersion relation. 

Regarding the magnetic compressibility $C_B$, our analytical analysis within the MHD-EBM framework reveals that plasma expansion plays a major role in shaping the radial evolution of magnetosonic fluctuations in the inner heliosphere, even in a linear regime. Both the expanding and non-expanding models show an increase in magnetic compressibility with radial distance, whereas the first-principles MHD-EBM treatment shows better agreement with in-situ observations. In particular, for fast solar wind conditions ($\xi = - 0.05$), choosing $R_0 = 0.1$ AU and $\theta = 5^\circ$, we obtain a radial profile of $C_B \propto R^{1.66}$, remarkably close to the observational scaling $C_B \propto R^{1.68\pm 0.23}$ reported by \citet{Chen_2020}. This agreement is significantly improved compared to the non-expanding case, suggesting that solar wind expansion plays a major role in controlling the radial evolution of magnetic compressibility~\citep{Chen_2020,Zhao_2021}.
These results are consistent with a gradual evolution of solar wind fluctuations from predominantly Alfvénic (transverse) at small heliocentric distances close to the Sun to more compressive at larger distances, around 1 AU. The inclusion of plasma expansion from the EBM formalism reveals a minimum in magnetic compressibility as the solar wind expands. The stabilization of this minimum into a more linear behavior is consistent with the transition between two turbulence regimes in the solar wind, an inner, Alfvénic region and an outer, more hydrodynamic region, at around $R \sim 0.3-0.4$ AU~\citep{Bruno2013,Chen_2020,Brodiano2023}. The agreement between the radial evolution of $C_B$ predicted by the MHD-EBM formalism for this set of parameters and the observed radial trends suggests that solar wind expansion is an active driver of the radial evolution of compressive turbulence throughout the heliosphere. However, confirming the existence and spatial distribution of this transition requires a more systematic comparison with in-situ measurements of both magnetic and velocity fluctuations in future studies.

On the other hand, the radial evolution of the ratio between the parallel electric and perpendicular magnetic field amplitudes, $|\delta E_\parallel|/|\delta B_\perp|$, reveals the distinct predictions of the non-expanding and expanding MHD frameworks. In both cases, the fast and slow magnetosonic modes respond in opposite ways to the radial decay of the Alfvén and sound speeds. As the fast-mode phase velocity increases relative to the Alfvén speed with distance, it becomes progressively more electrostatic, whereas the slow-mode phase velocity decreases, shifting it toward a more magnetically dominated regime. This behavior can be traced to expansion-induced anisotropies in the background magnetic field, which can enhance or suppress the parallel electric field component associated with each mode. Beyond a radial distance of $R\sim 0.17$ AU for the parameter set considered here, the expanding and non-expanding cases diverge significantly. While the non-expanding case predicts a continued radial growth of $|\delta E_\parallel|/|\delta B_\perp|$ for both magnetosonic modes, the expanding case shows a reduced radial slope, suppressing the increase of the ratio with radial distance. This reflects the effect of plasma expansion on the relation between the electric and magnetic field components of the fluctuations. Together with the radial evolution of the magnetic compressibility, these results suggest that solar-wind expansion plays a major role in modifying the balance between electrostatic and electromagnetic contributions of magnetosonic fluctuations in the inner heliosphere, in a way that is not captured by non-expanding radial rescaling models.

Further applications of the theoretical framework used and the dispersive properties radial profiles deduced here can be made. The analytical results and approach presented in this work provide a solid basis for future studies of expanding plasma turbulence and wave propagation. In particular, this approach can be extended to investigate the nonlinear coupling between MHD-EBM modes, the role of multi-fluid effects, and the evolution of rugged invariants in expanding turbulence. Such extensions will enable a more comprehensive understanding of the effects of radial solar wind expansion and background magnetic field structure on the energy cascade and dissipation across scales.

\begin{acknowledgments}
    We are grateful for the support of ANID Chile through the National Doctoral Scholarships No. 21250323 (SS), and FONDECYT grants No. 1240281 (PSM) and 1230094 (FAA). 
\end{acknowledgments}

\begin{appendix}

\section{Radial dependence in non-expanding dispersion relation}
\label{appendixa}
Here, we derive the dispersion relation of MHD waves in a non-expanding frame and subsequently apply the radial scales of the plasma quantities quantities after the derivation, in order to assess the consistency of this approximation and compare with our derivation from first principles within the EBM formalism. 

In the standard MHD frame, assuming $\mathbf k = (k,0,0)$ and $\mathbf B_0 = B_{0x}(1,\xi',\xi')$, where $\xi'$ is the non-expanding magnetic spiraling ratio, the elements of the resulting dispersion matrix are
\begin{align}
    D_{xx}& = \omega^2/k^2 -  2  v_A^2 \xi'  -  v_S^2 ,\\
    D_{xy}& =D_{xz} =      D_{yx}  =  D_{zx} =   v_A^2 \xi'  ,\\
    D_{yy} & = D_{zz} = \omega^2/k^2 - v_A^2,\\
    D_{yz} & = D_{zy} = 0.
\end{align}
The corresponding eigenfrequencies in the non-expanding frame are
\begin{equation}
    \begin{split}
        \omega^2 & = \frac{k^2}{2} ( v_A^2(1 + 2 \xi^2) +  v_S^2  + [(v_S^2 + v_A^2 (1 + 2 \xi^2 ))^2  \\&- 4 v_A^2 v_S^2 ]^{1/2}) 
    \end{split}
\end{equation}
\begin{equation}
    \begin{split}
        \omega^2 & = \frac{k^2}{2} ( v_A^2(1 + 2 \xi^2) +  v_S^2  - [(v_S^2 + v_A^2 (1 + 2 \xi^2 ))^2  \\&- 4 v_A^2 v_S^2 ]^{1/2}) 
    \end{split}
\end{equation}
\begin{equation}
    \omega^2 = k^2 v_A^2. 
\end{equation}
Note that these expressions correspond to the non-expanding limit $(a = 1)$ of Equations \eqref{fast}-\eqref{alfven}. In order to mimic the effect of plasma expansion, we introduce radial profiles for the background quantities: $\tilde v_A \propto v_A/R$,  $\tilde v_S \propto v_S/R^{\gamma - 1}$, $\mathbf B_0 \propto (1/R^2, 1/R,1/R)$ thus defining the expanding $\xi$ parameter as $\xi = a \xi'$. Substituting these profiles in the above dispersion relation yields
\begin{equation}
    \begin{split}
        \omega^2 & = \frac{k^2}{2} (\tilde v_A^2(1 + 2 a^2 \xi^2) +  \tilde v_S^2  + [(\tilde v_S^2 + \tilde v_A^2 (1 + 2 a^2 \xi^2 ))^2  \\&- 4 \tilde v_A^2 \tilde v_S^2 ]^{1/2}) \label{disp1}
    \end{split}
\end{equation}
\begin{equation}
    \begin{split}
        \omega^2 & = \frac{k^2}{2} (\tilde v_A^2(1 + 2 a^2 \xi^2) +  \tilde v_S^2  - [(\tilde v_S^2 + \tilde v_A^2 (1 + 2 a^2 \xi^2 ))^2  \\&- 4 \tilde v_A^2 \tilde v_S^2 ]^{1/2}) \label{disp2}
    \end{split}
\end{equation}
\begin{equation}
    \omega^2 = k^2 \tilde v_A^2. \label{disp3}
\end{equation}

The comparison of these expressions to Equations \eqref{fast}-\eqref{alfven} shows that, while the Alfvén mode dispersion relation can be recovered by assuming a radial dependence of Alfvén speeds, the dispersion relation for magnetosonic modes derived from first principles and assuming a radial profile of  plasma speeds are not equivalent.

%\section{Polarization for a non-uniform field}
%\label{appendixb}
%Polarization relative to the z-axis was defined by ~\cite{Stix1992} as 

%\begin{equation}
%    P = i \frac{\delta E_x}{\delta E_y} = i \left ( \frac{D_{xz}D_{yy} - D_{xy}D_{yz}  }{D_{xx}D_{yz} - D_{xz}D_{yx}  }\right ).\label{eqpol}
%\end{equation}

%Where $D_{ij}$ are the components of the dispersion tensor in each case. Nevertheless, to define polarization relative to the background magnetic field in this configuration, it is necessary to write a basis $\{\hat e_1, \hat e_2, \hat B_0 \}$, where $\hat e_1$ and $\hat e_2$ are unitary vectors orthogonal to the unitary vector $\hat B_0 = \mathbf B_0/B_0$. This will allow the transformation to the new basis of the dispersion tensor, as $\mathbb D' = \mathbf R \mathbb D \mathbf R^T$, where $\mathbf R$ are the transformation matrices built from the orthogonal basis. We define the orthogonal basis as
%\begin{align}
%     & \hat e_1 = \frac{1}{\sqrt{2}}(0,1,-1), \\ & \hat e_2 = \frac{1}{\sqrt{2}} \frac{1}{\sqrt{1 + 2 a^2 \xi^2}} (- \sqrt{2 a \xi}, 1, 1).
%\end{align}
%Redefining the dispersion tensor in this new basis, we obtain the polarization with respect to the background magnetic field as
%\begin{equation}
    %P = i \frac{\delta E_{\hat e_1}}{\delta E_{\hat e_2}} = \frac{D'_{\hat e_1 \hat B_0}   D'_{\hat e_2 \hat e_2} - D'_{\hat e_1 \hat e_2} D'_{\hat e_2 \hat B_0}}{D'_{\hat e_1 \hat e_1} D'_{\hat e_2 \hat B_0} - D'_{\hat e_1 \hat B_0}  D'_{\hat e_1 \hat B_0}}.
%\end{equation}

\section{B-Fields}
\label{appendixc}
In order to calculate the magnetic field perturbation eigenvector $\delta \mathbf B$, the expanding Faraday's Law given by \eqref{fara} can be linearized using previous principles. For the first-order equation, we obtain 
 \begin{equation}
    (\mathbb A^{-1} \cdot \nabla) \times \delta \mathbf E = - \frac{1}{c} \left ( \frac{\partial \delta \mathbf B}{\partial t} + \frac{\dot a}{a} \mathbb L \cdot \delta \mathbf B\right ).
\end{equation}
Multiplying both sides by $\mathbb Z = \text{diag}(a^2,a,a)$, a product rule appears at the right-hand side of the equation. Thus, the linearized expanding Faraday's law can be written as
\begin{equation}
    \mathbb Z \cdot (\mathbb A^{-1} \cdot \nabla) \times \delta \mathbf E = - \frac{1}{c} \frac{\partial}{\partial t} \left ( \mathbb Z \cdot\delta  \mathbf B\right ).
\end{equation} 

This equation can be solved through standard Fourier analysis, where the operators $\nabla$ and $\partial/\partial t$ are replaced by $i\mathbf{k}$ and $-i\omega$, respectively. Under solar wind conditions, the $1/\tau$ frequency is much smaller than the wave frequency $\omega$, being about five orders of magnitude smaller than the proton gyrofrequency at 1 AU~\citep{Bruno2013}. Hence, the $\mathbb Z$ term can be considered slowly varying and thus neglected in the Fourier transform. Therefore, MHD wave oscillations evolve much faster than the expansion, allowing us to write the $\delta \mathbf B$ components as
\begin{align}
    \delta B_x &= - \frac{k c}{\omega a^2} \delta E_y \sin \theta,\label{eqbx} \\
    \delta B_y &= \frac{kc}{\omega a} \left ( \delta E_x \sin \theta - a^2 \delta E_z \cos \theta\right ),\label{eqby} \\
    \delta B_z & = \frac{k c a}{\omega } \delta E_y \cos \theta. \label{eqbz} 
\end{align}
Here, we have assumed a wavevector of the form $\mathbf{k} = k(\cos \theta, 0, \sin \theta)$, to quantify following properties as functions of the $\theta$ angle.
Under this configuration, the polarization properties of the modes can be directly assumed. For the Alfvén mode, $\delta B_z = 0$, $\delta B_x = 0$, and $\delta B_y \neq 0$, while for the two magnetosonic modes, $\delta B_y = 0$ and $\delta B_x, \delta B_z \neq 0$. Thus, for the Alfvén mode, the perturbations are purely transverse to both the background magnetic field and the propagation direction, as the oscillations in $\delta B_y$ are associated with shear Alfvén waves. On the other hand, the magnetosonic modes are associated with compressional perturbations within the plane defined by the background magnetic field and the wavevector, thus manifesting as fluctuations in $\delta B_x$ and $\delta B_z$.

\section{Electric-to-magnetic amplitude ratio}
\label{appendixd}
This quantity can be written as
\begin{equation}
    \frac{|\delta E_\parallel|}{|\delta B_\perp|} = \frac{|\delta \mathbf E \cdot \hat{ \mathbf B_0}|}{\sqrt{\delta \mathbf B^2 - \delta \mathbf B_\parallel^2 } }\label{defratio}.
\end{equation}
Where we have used $\delta \mathbf B^2 = \delta \mathbf B_\perp^2 + \delta \mathbf B_\parallel^2$. First, we write the parallel electric field component as
\begin{align}
    \delta \mathbf E_\parallel = \frac{|\delta E_x + \delta E_y a \xi + \delta E_z a \xi|}{\sqrt{1 + 2 a^2 \xi^2 }}.\label{deltae}
\end{align}
On the other hand, the perpendicular component of the magnetic field is
\begin{equation}
    \begin{split}
            |\delta \mathbf B_\perp| &=  \sqrt{\delta\mathbf B^2 - \delta \mathbf  B_\parallel^2 } \\ &= \Bigg  [\delta B_x^2 + \delta B_y^2 + \delta B_z^2 \\ & - \left .\frac{(\delta B_x + a \xi (\delta B_y + \delta B_z))^2}{1 + 2 a^2 \xi^2}\label{deltab}\right ]^{1/2}.
    \end{split}
\end{equation}

Using Equations \eqref{eqbx}-\eqref{eqbz} in Equation \eqref{deltab}, and then the resulting Equation \eqref{deltab}, \eqref{deltae} in Equation \eqref{defratio}, we obtain
\begin{equation}
    \begin{split}
        \frac{|\delta E_\parallel|}{|\delta B_\perp|} = \frac{\omega}{kc} \frac{|\delta E_x + a \xi (\delta E_y + \delta E_z)|}{\displaystyle{\sqrt{K(a,\xi)}}},
    \end{split}
\end{equation}
where
\begin{equation}
    K(a,\xi) = Q - \frac{S^2}{1 + 2 a^2 \xi^2}.\label{K}
\end{equation}
Here,
\begin{equation}
    S = \xi \delta E_x \sin \theta + \delta E_y \left (a^2 \xi \cos \theta - \frac{\sin^2 \theta}{a^2} \right ) - a^2 \xi \delta E_z \cos \theta,\label{s}
\end{equation}
and
\begin{equation}
    \begin{split}
        Q = \delta E_y \left ( \frac{\sin^2 \theta}{a^4} + a^2 \cos^2 \theta\right ) + \delta E_x^2 \frac{\sin^2}{a^2} \\ + a^2 \delta E_z^2 \cos^2 \theta - 2 \delta E_x  \delta E_z \sin \theta \cos \theta.\label{q}
    \end{split}
\end{equation}

\end{appendix}

\bibliography{references}{}
\bibliographystyle{aasjournalv7}



\end{document}